\documentclass[showpacs,preprintnumbers,amsmath,amssymb,superscriptaddress]{revtex4}
\usepackage{epsfig}
\usepackage{graphicx}
\usepackage{dcolumn}
\usepackage{bm}

\begin{document}

\preprint{CAB-lcasat/04005}
\title{Dynamic Renormalization Group and Noise Induced Transitions in a Reaction
  Diffusion Model.}


\author{M.P. Zorzano}
\affiliation{Centro de Astrobiolog\'{\i}a (CSIC-INTA),
        Carretera de Ajalvir km 4,    28850 Torrej\'{o}n de Ardoz, Madrid, Spain}   

\email{zorzanomm@inta.es}

\homepage[url]{www.cab.inta.es}

\author{David Hochberg}
\affiliation{Centro de Astrobiolog\'{\i}a (CSIC-INTA),
        Carretera de Ajalvir km 4,    28850 Torrej\'{o}n de Ardoz,
	Madrid, Spain}
\author{Federico Mor\'{a}n}
 \affiliation{Centro de Astrobiolog\'{\i}a (CSIC-INTA),
        Carretera de Ajalvir km 4,    28850 Torrej\'{o}n de Ardoz,
	Madrid, Spain}
\affiliation{Departamento de Bioqu\'{\i}mica y Biolog\'{\i}a Molecular, Facultad de
Ciencias Qu\'{\i}micas, Universidad Complutense de Madrid, Spain}

\date{\today}

\begin{abstract}
We investigate how additive weak noise (correlated as well as
uncorrelated) modifies the parameters of the Gray-Scott (GS)
reaction diffusion system by performing numerical simulations and
applying a Renormalization Group (RG) analysis in the neighborhood
of the spatial scale where biochemical reactions take place. One
can obtain the same sequence of spatial-temporal patterns by means
of two equivalent routes: (i) by increasing only the noise
intensity and keeping all other model parameters fixed, or (ii)
keeping the noise fixed, and adjusting certain model parameters to
their running scale-dependent values as predicted by the RG. This
explicit demonstration validates the dynamic RG transformation for
finite scales in a two-dimensional stochastic model and provides
further physical insight into the coarse-graining analysis
proposed by this scheme.  Through several study cases we explore the
  role of noise and its temporal correlation in self-organization and propose a way to drive the system
  into a new desired state in a controlled way.
\end{abstract}


\maketitle
\section{Introduction.}

The dynamics of certain biological systems frequently follows some
self-organization process where the development of new, complex
structures takes place primarily in and through the system itself.
This {\sl self-organization} is normally triggered by internal variation
processes, which are usually called "fluctuations" or "noise" , that have a
positive influence on the system.
 For instance, recent theoretical studies and experiments with
  cultured glial cells and the Belousov-Zhabotinsky reaction have shown that
  noise may play a constructive role on the dynamical behavior of spatially
  extended systems \cite{Jung,Sancho,Showalter}. Therefore, one can not ignore the role of noise in chemical and
biological self-organization and its relationship with the
environmental selection of emergent patterns \cite{SO}.

  In these systems there is a strong interplay between the equations
of the microscopic dynamics and the environmental fluctuations
acting on the smallest scales. In order to get a qualitative idea
of the system behavior one can average out the influence of noise
on the shortest scales by increasing,  infinitesimally, the scale
of observation. This can be done applying dynamic
Renormalization Group (RG) techniques \cite{Ma,Cardy}. One obtains
then the so-called RG flow equations which are generally used to
study the RG flow of the system parameters and to find the fixed
points in the long wavelength limit (at the fixed points the
system is invariant under the scaling transformation). But our
interest in this paper is in the finite-size renormalization (at
finite wavelengths or finite scales), i.e. on the influence of
noise at the scale where the chemical or biological reactions take
place. 

 Our final aim is to validate
 the RG procedure as a general tool which can be used at the scales where
 these biochemical reactions take place and predict the
 transitions induced by noise with different correlation properties. As an example we will investigate the
influence of noise on a reaction diffusion system in the vicinity
of a transition point. Performing numerical simulations and applying the RG analysis at
finite scale ranges we can get a better understanding of how weak noise modifies the
parameters of this  dynamical system and the patterns it converges to.

We consider the Gray-Scott (GS) reaction
diffusion model \cite{GS}, which is one
 of the simplest models of biochemical relevance leading to spatial and
temporal patterns when diffusion is included.
Numerical simulations of this system in the \textit{deterministic} case have revealed a
surprisingly large set of complex and irregular patterns as a function of the
system parameters
\cite{Pearson}. In addition we want to include the influence of the
fluctuating environment. Fluctuations of {\sl internal} (thermal) origin
represent microscopic degrees of freedom, and since they evolve in spatial
and temporal scales much shorter than those of the gross variables of the
system they are assumed to be additive and uncorrelated in space and time. When the origin of
the noise is {\sl external} there may exist a coupling between the system and
the fluctuations. In this case there is no difference between the time and
length scales of the noise and the field, and this noise could be
multiplicative and have some structure in space or time. We study the
following stochastic system with additive noise:
\begin{eqnarray}\label{GSN}
\frac{\partial V}{\partial t}&=&\lambda U V ^{2} -\mu V + D_v \nabla ^{2} V
+\eta_v (x,y,t)\\ \nonumber
\frac{\partial U}{\partial t}&=&u_0-\lambda  U V ^{2} -\nu U + D_u \nabla ^{2} U
+\eta_u (x,y,t)
\end{eqnarray}
where $\nabla ^{2}=\frac{\partial ^{2}}{\partial x^{2}}+\frac{\partial
  ^{2}}{\partial y^{2}}$. We will consider Gaussian white zero-mean noises (uncorrelated)
\mbox{$<\eta _{u,v}(x,y,t)\eta _{u,v}(x',y',t')>$}$=2
A_{u,v}$\mbox{$\delta(x-x')$}\mbox{$\delta(y-y')$}\mbox{$\delta(t-t')$}
and Ornstein-Uhlenbeck (OU) correlated noise $<\eta
_{u,v}(x,y,t)\eta _{u,v}(x',y',t')>= \hat{A}_{u,v}
\exp{\frac{-|t-t'|}{\tau}}\delta(x-x')\delta(y-y')$ (the Gaussian
white noise case limit  $2A_{u,v}\delta(t-t')$  is obtained when
the correlation time $\tau \rightarrow 0$ while keeping
${A}_{u,v}= \hat{A}_{u,v}\tau$ constant). In the absence of noise,
Eq. \ref{GSN} defines the Gray-Scott model, which is a variant of
the autocatalytic Selkov model of glycolisis, corresponding to the
following chemical reactions:

\begin{eqnarray}\label{GS2}
U +2V & \stackrel{\lambda}{\rightarrow}& 3 V \\ \nonumber
V &\stackrel{\mu}{\rightarrow}&  P\\\nonumber
U  &\stackrel{\nu}{\rightarrow}& Q \\ \nonumber
&\stackrel{u_0}{\longrightarrow}&  U.\\\nonumber
\end{eqnarray}
$V(x,y,t)$ and $U(x,y,t)$ represent the concentrations of the
chemical species $U$ and $V$, and are functions of the
two-dimensional space and time variables. 
$\lambda$ is the constant rate of the reactions, $P$ and $Q$ are inert
products, $\mu$ is the decay
rate of $V$ and $\nu$ the decay rate of $U$. The equilibrium
concentration of $U$ is $u_0/\nu$, where $u_0$ is the feed rate
constant. The chemical species $U$ and $V$ can diffuse with
independent diffusion constants $D_u$ and $D_v$. All the model
parameters, including the noise amplitudes $A_{u,v}$, are
positive.

In the case of Gaussian white noise the stochastic system
described in Eq. \ref{GSN} has been numerically integrated for a
particular set of parameters \cite{FL}. The pattern to which the
system converged changed drastically with small changes in the
noise intensity, suggesting that this system can suffer noise
induced transitions when subjected to random fluctuations
\cite{NIT}.

Additionally, one can apply the RG transformation to the
Gray-Scott model subject to additive random fluctuations
\cite{DH}. The dynamic RG procedure consists of two steps: an
elimination or thinning out of the fast or short wave-length modes
($\frac{\Lambda}{s} < |\vec{k}| < \Lambda$ where $s=e ^{l} \geq 1$
is the scale in wavenumber space above which the fluctuations have
been integrated over, $\Lambda=\frac{2\pi}{L}$ and $L$ the minimal
length where the equations under consideration are valid) followed
by a rescaling of the remaining modes ($\vec{k}\rightarrow
s\vec{k}$) \cite{Ma}. Applying this infinitesimal transformation
one gets the effective (i.e. renormalized) parameters of the
system when going through an infinitesimal scale change. This
analysis leads one to consider a set of coupled differential
equations governing the RG flow in parameter space when this
system is influenced by noise. At one-loop order in the noise
amplitude, only two out of the total of eight model parameters run
with scale, namely, the decay rate $\nu$, of the $U$ field and the
nonlinear coupling $\lambda$, and the corrections are driven by
the $v$-noise (not the $u$-noise). In particular for the case of
white Gaussian noise, in a two-dimensional space, the pertinent
equations read \cite{DH}
\begin{eqnarray}\label{RG_para}
\frac{d \nu}{d l}&=&z \nu + \frac{\lambda A_v K_2 \Lambda ^{2}}{(\mu+ D_v
  \Lambda ^{2})}\\ \nonumber
\frac{d \lambda}{d l}&=& \Big(2\chi + z - \frac{4\lambda A_v K_2
\Lambda ^{2}}{(\mu+
  D_v \Lambda ^{2})(\mu+\nu +D_v \Lambda ^{2}+D_u \Lambda ^{2})} \Big)\lambda \nonumber
\end{eqnarray}
 and for the case of OU noise with long-range correlations read:

\begin{eqnarray}\label{RG_OU}
\frac{d \nu}{d l}&=&z \nu +
\csc{\left[\frac{(1+2\theta_v)\pi}{2}\right]}\frac{\lambda {A}_v K_2 \Lambda^{2}}{(\mu+ D_v \Lambda ^{2})^{1+2\theta_v}}\\ \nonumber
\frac{d \lambda}{d l}&=&(2\chi + z )\lambda\\ \nonumber
&\ &-4 \csc{\left[\frac{(1-2\theta_v)\pi}{2}\right]}\ \lambda {A}_v K_2
  \Lambda^{2} (\nu+ D_v \Lambda ^{2})  \\ \nonumber
&\ &\frac{(\nu+ D_u \Lambda
    ^{2})^{-1-2\theta_v}-(\mu+ D_v \Lambda ^{2})^{-1-2\theta_v}}{(\mu+ D_v
    \Lambda ^{2}) ^{2}-(\nu+ D_u \Lambda ^{2}) ^{2}}\lambda
\end{eqnarray}
 with $A_v=\hat{A}_v\tau$ and $\theta_v=1$ \footnote{These equations have been
  derived in \cite{DH} for a general case of noise with memory 
  and a power law spectral density $S(w)=A_v
{w}^{-2 \theta_v}$. In particular, for the OU noise with  spectral density 
$S(w)=\frac{\hat{A}_{u,v}\tau }{1+\omega ^{2} \tau ^{2}}$, the value of the decay exponent
  $\theta_v$ 
  is obtained in the limit of long correlation time $\tau \rightarrow \infty$
  (but such that $A_v=\hat{A}_v\tau$ is, again,
kept constant) rescaling the time to $\hat{t}=t/\tau$ and therefore rescaling the
frequency to $\hat{w}=w\tau$ 
 so that $S(\hat{w})=A_v
\hat{w}^{-2}$. }. Here $d l$ is an infinitesimal change in
the scale $s$ of the system (from $s=e^{l}$ to $\hat{s}=s+ds$ with
$ds=sdl$). The scaling parameters $z$ and $\chi$ are the so called
dynamic exponent (for time) and roughness exponent (for the fields
$U$ and $V$) and are not needed for the present study.

These equations (\ref{RG_para},\ref{RG_OU}) predict how the
effective parameters of the system run with scale and noise
amplitude and have been derived and analyzed in detail for the
long-wavelength limit (i.e. in the large scale limit $s\rightarrow
\infty$) in \cite{DH}. Our aim in this paper is to validate these
equations in the intermediate scale range, far from the fixed
points, where the chemical processes take place. To carry this
out, we will use (\ref{RG_para},\ref{RG_OU}) to calculate the
effective parameters of the GS system for a finite $\Delta l$ and
noise of significant intensity. We then solve (\ref{GSN}) using
these effective parameters (and negligible noise), and compare the
resultant patterns based on these running parameters with those
that are obtained numerically
when all the model parameters are held fixed and only the noise intensity
  is increased.

\section{Dynamic RG theory with finite scaling and uncorrelated noise.}

The RG consists of thinning the degrees of freedom followed by a
re-scaling of length and time \cite{Ma}. It is a general
calculational \textit{scheme} for treating problems where
fluctuations at many length scales are important.  The final
\textit{aim} of the RG is to describe how the dynamics of a system
subject to random fluctuations evolves as one changes the scale of
observation. In practice, the  RG procedure permits one to
establish rigorous correspondences between sets of parameters
defining physically different states observed at different scales.
Indeed, (\ref{RG_para},\ref{RG_OU}) express the fact that $\nu$ and
$\lambda$ change with the scale $l$, as functions of the noise and the
other model parameters, as a consequence of the coarse-graining.
In particular, in the case
under consideration, this correspondence is due to the presence of
noise and a finite change in the spatial scale.

Following Eq. (\ref{RG_para}) we observe, that for a scale
variation of $\Delta l$ the predicted running or modification of
the $\nu$ and $\lambda$ parameters induced by noise alone, should
be $\Delta \nu= + \Delta l \frac{\lambda A_v K_2
\Lambda^{2}}{(\mu+ D_v \Lambda ^{2})}$ and $\Delta \lambda= -
\Delta l \ \lambda \left(\frac{4\lambda A_v K_2 \Lambda
^{2}}{(\mu+ D_v \Lambda ^{2})(\mu+\nu +D_v \Lambda ^{2}+D_u
\Lambda ^{2})}\right)$ respectively. For the two-dimensional case,
$K_2=1/(2\pi)$. In our numerical simulation the smallest scale of
observation is that of the cell with $\Delta x \times \Delta y$
and $\Delta x=\Delta y$. This determines the minimal wavelength
and maximal momentum $\Lambda=\frac{2\pi}{\Delta x}$ that can be
considered in the problem. Let us take the cell as our initial
scale, our scaling factor is therefore $s=1$.  The diffusion term
allows us to average data over neighboring cells, this in turn
defines a greater scale of approximately $3 \Delta x \times 3
\Delta y$ which leads us to $\hat{s}=s+\Delta s$ with $\Delta
s=2=s \Delta l$ and $\Delta l=2$.

We will explore the patterns that are formed within a certain
range of the parameter space. Similar studies have been performed
in the deterministic Gray-Scott model, keeping $\lambda =1$
constant, varying $\mu$ between $0.06$ and $0.14$ and $\nu$
between $0.01$ and $0.07$ (in this region there is a transition
from two stable steady states to one trivial state and a great
variety of patterns) \cite{Pearson}.

In Table \ref{table1} we show, for a given noise intensity $A_v$,
the predicted effective parameters $\lambda+\Delta \lambda$ and
$\nu+\Delta \nu$ based on the one-loop RG calculations in
\cite{DH}, for $\lambda=1,D_v=0.5,D_u=1$, $u_0=0.05=\nu$ and
$\mu=0.1155$. The labels in column 1 and 5 will be used later to name
the study cases. In Table \ref{table2} we show the same
calculations, in a different region of parameter space, when the
unperturbed values are $\nu=u_0=0.03$ and $\mu=0.086$. In both
cases we have set $\nu=u_0$ so that the trivial state solution is
always $U=1$.

\begin{table}[htb]
\begin{center}
\caption{\label{table1}Effective or renormalized parameters (sixth
and seventh columns) induced by a white noise of intensity $A_v$
when the unperturbed or bare values are $\mu=0.1155$,
$\nu=u_0=0.05$ and $\lambda=1$.}\vskip 1.5mm
\begin{tabular}{|c|c|c|c||c|c|c|}    \hline\hline
 case & $A_v$ & $\lambda$ & $\nu$ &case & $\lambda+\Delta \lambda$ & $\nu+\Delta \nu$  \\ \hline
A  &  $4.8\times 10^{-6}$ & $1$ & $0.05$ & a &$1$ (-$10 ^{-6}$) &  $0.05$ (+$3 \times 10 ^{-6}$) \\ \hline
 B &$1.61\times 10^{-4}$ &  $1$ & $0.05$  & b & $0.99997$ &  $0.0501$\\ \hline
 C & $4.36\times 10^{-4}$ &  $1$ & $0.05$  & c & $0.99991$ &  $0.05027$ \\ \hline
 D &$4.84\times 10^{-4}$ &   $1$ & $0.05$  & d &  $0.999903$ &  $0.0503$  \\ \hline\hline
\end{tabular}
\end{center}
\end{table}
\begin{table}[htb]
\begin{center}
\caption{\label{table2}Effective or renormalized parameters (sixth
and seventh columns) induced by a white noise of intensity $A_v$
when the unperturbed or bare values are $\mu=0.086$,
$\nu=u_0=0.03$ and $\lambda=1$.}\vskip 1.5mm
\begin{tabular}{|c|c|c|c||c|c|c|}    \hline\hline
case & $A_v$ & $\lambda$ & $\nu$ & case & $\lambda+\Delta \lambda$ & $\nu+\Delta \nu$  \\
\hline
I &  $4.8\times 10^{-6}$ &  $1$ & $0.03$  &  i &$1$ (-$ 10 ^{-6}$) &  $0.03$ (+$3  \times10 ^{-6}$) \\ \hline
II & $1.60\times 10^{-4}$ &   $1$ & $0.03$  & ii & $0.99997$ &  $0.0301$  \\ \hline
III  & $4.33\times 10^{-4}$ &  $1$ & $0.03$  & iii & $0.999914$ &  $0.03026$ \\ \hline
\hline 
\end{tabular}
\end{center}
\end{table}

The numerical simulations of system evolution have been performed
using forward Euler integration of the finite-difference equations
following discretization of space and time in the stochastic
partial differential equations. The spatial mesh consists of a
lattice of $256\times 256$ cells of size $\Delta x=\Delta y=2.2$,
with periodic boundary conditions. Noise has been discretized as
well. The initial conditions consisted of one localized square
pulse with ($U=0.5,V=0.25$) perturbing the trivial steady state
($U=1,V=0$) plus random Gaussian noise. The perturbing pulse
measured $22 \times 22$, just wide enough to allow the
autocatalytic reaction to be locally self-sustaining. The system
has been numerically integrated for up to 30000 time steps (step
size $\Delta t=1$). In the figures, only the concentration of the
substrate $U$ is shown.  When displayed in color, the blue
represents a concentration between 0.2 and 0.4, where the
substrate is being depleted by the autocatalytic production of
$V$, yellow represents an intermediate concentration of roughly
0.8 and red represents the trivial steady state ($U=1$).

\subsection{Noise Induced Transition}

\begin{figure}
\begin{tabular}{ccccc}
    (A) & (B) & (C) & (D)  \\
\includegraphics[width=0.24 \textwidth]{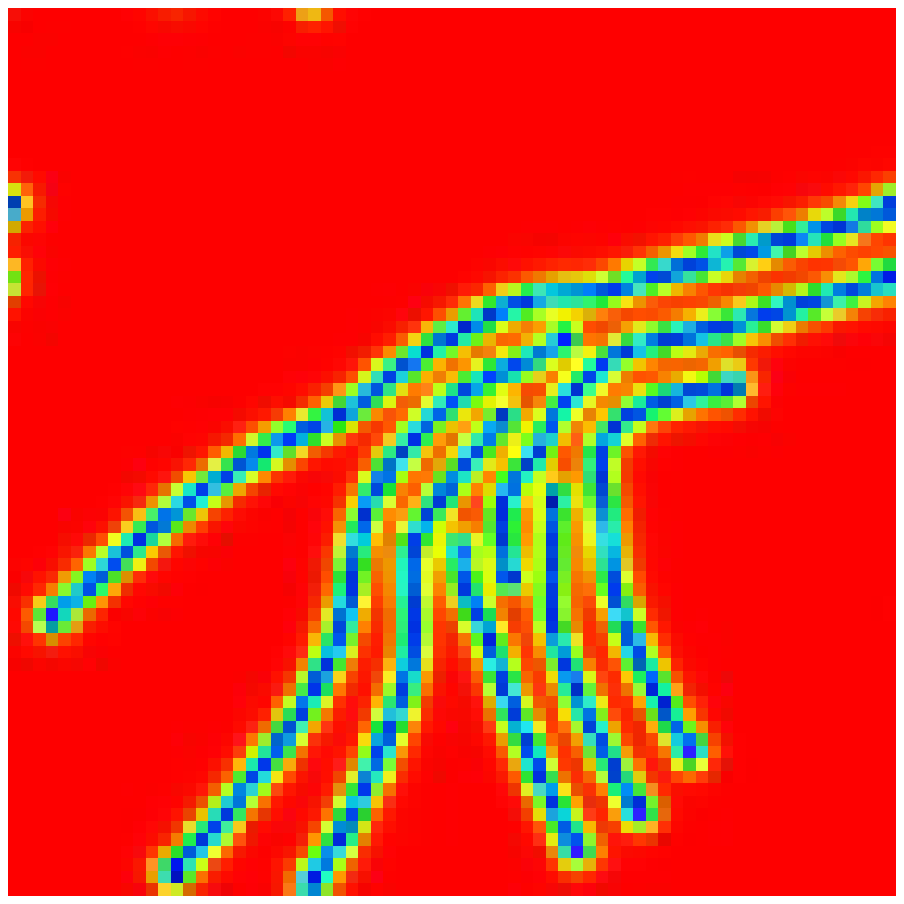} &
\includegraphics[width=0.24 \textwidth]{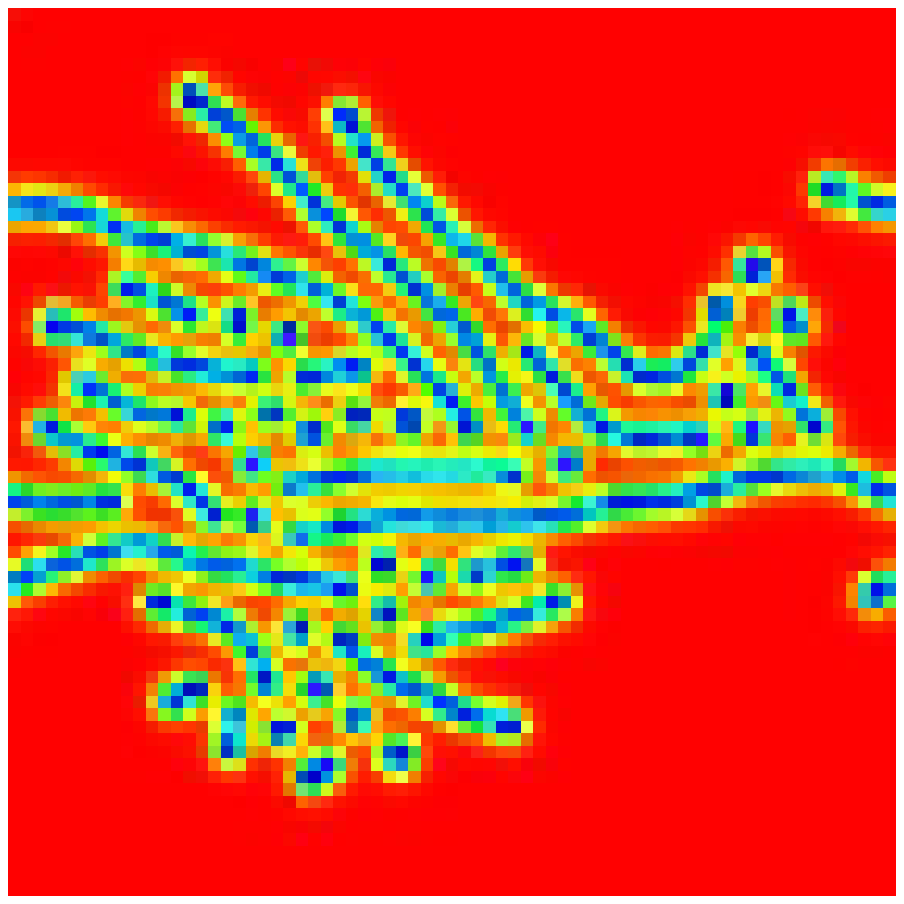} &
\includegraphics[width=0.24 \textwidth]{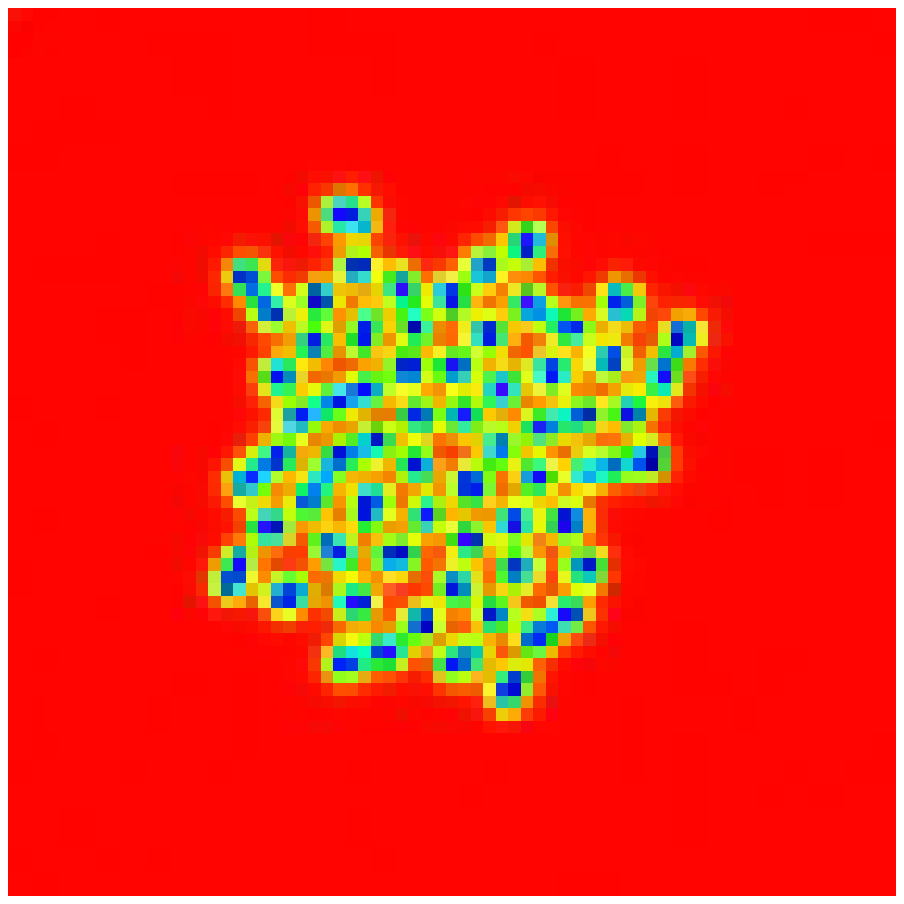} &
\includegraphics[width=0.24 \textwidth]{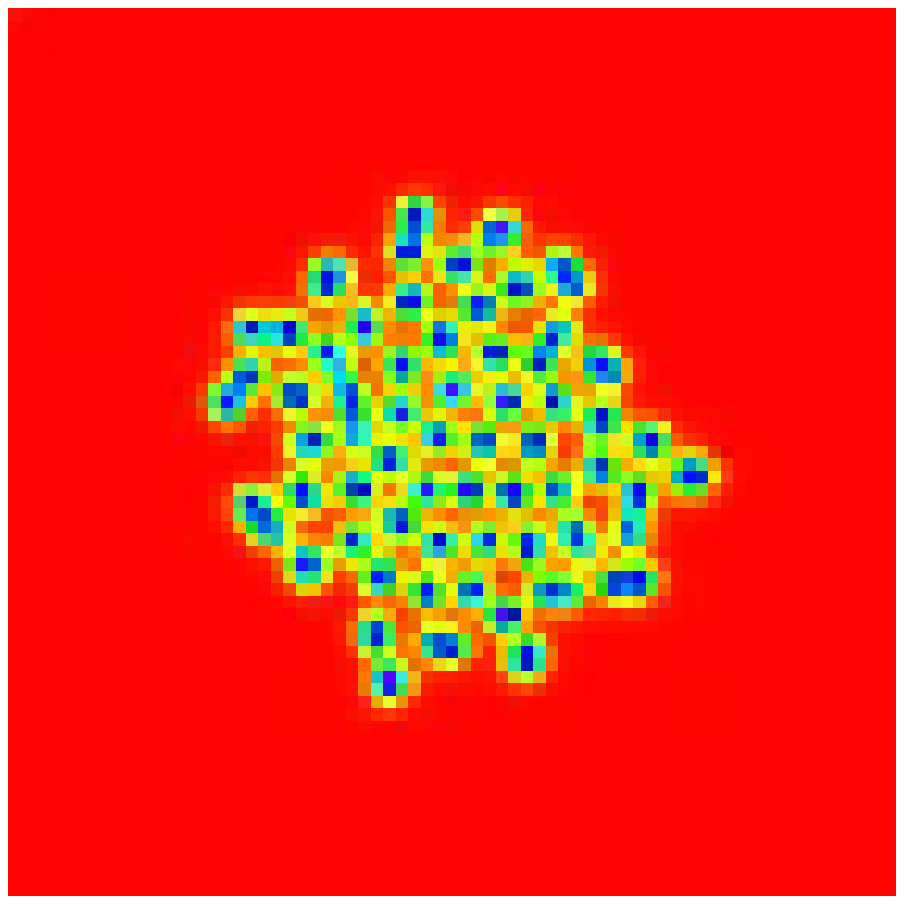}
\end{tabular}
\caption{\label{fig1} Results obtained with all parameters held
fixed and increasing only the noise intensity $A_v$ with the
values given in column two of Table \ref{table1}. (A) and (B) are
shown at $t=20000$, (C)-(D) at  $t=10000$. }
\begin{tabular}{cccc}
    (a) & (b) & (c) & (d)  \\
\includegraphics[width=0.24 \textwidth]{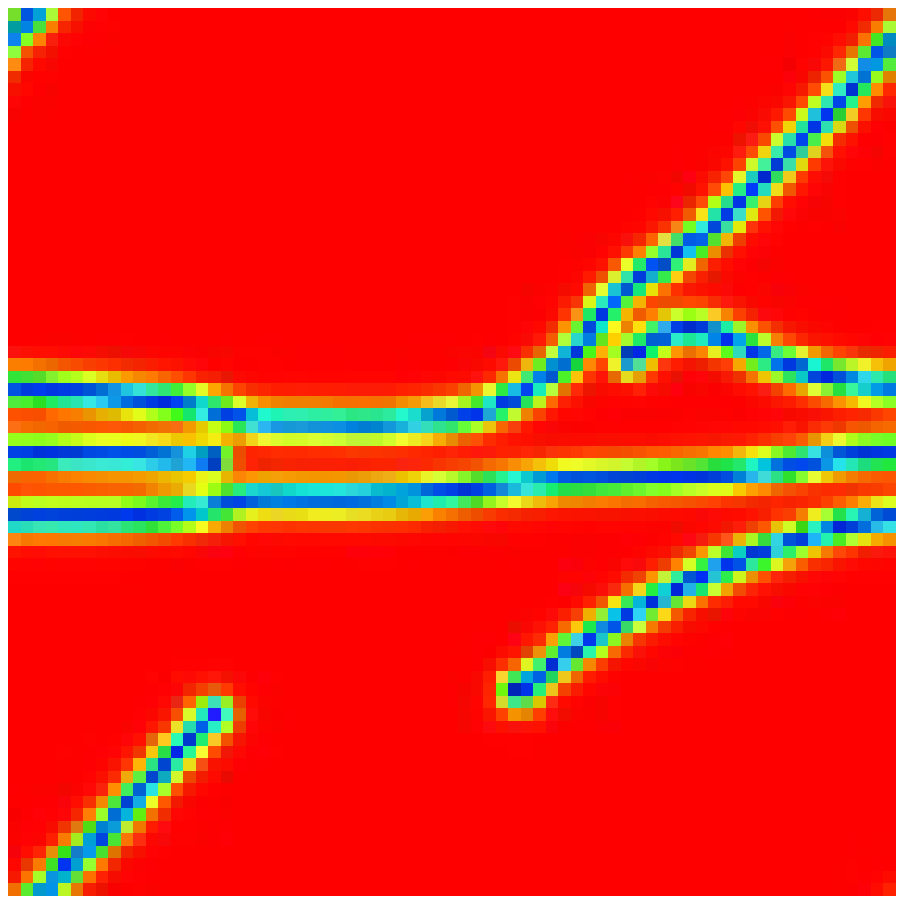} &
\includegraphics[width=0.24 \textwidth]{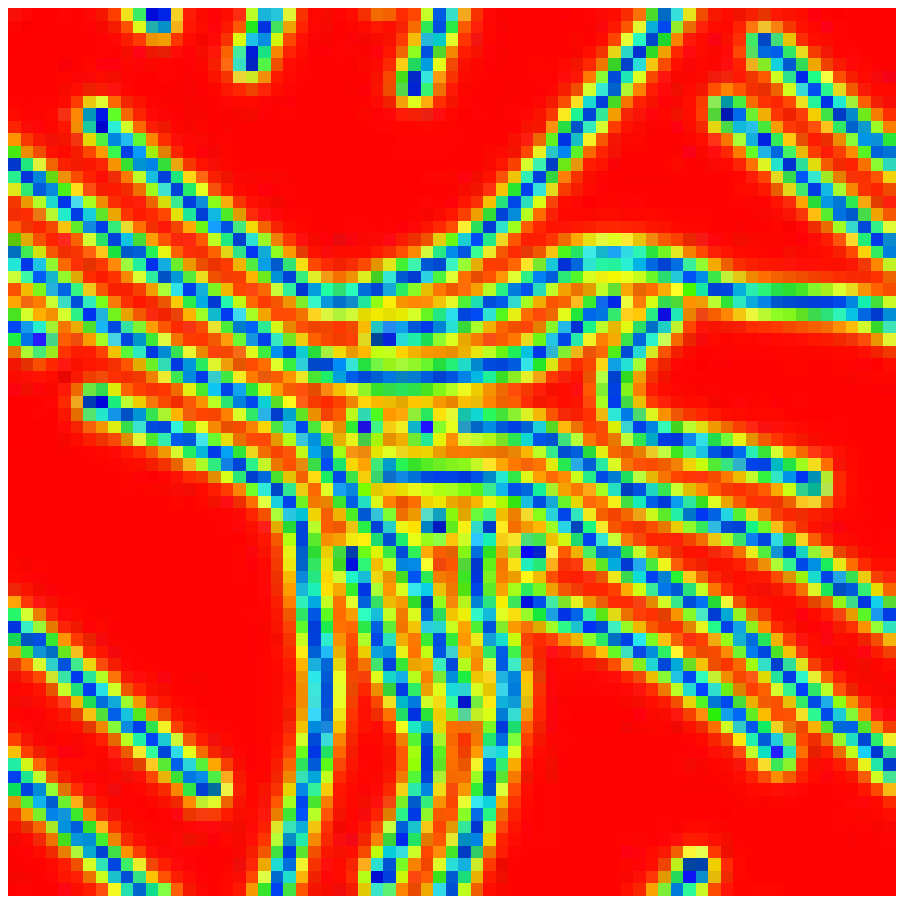} &
\includegraphics[width=0.24 \textwidth]{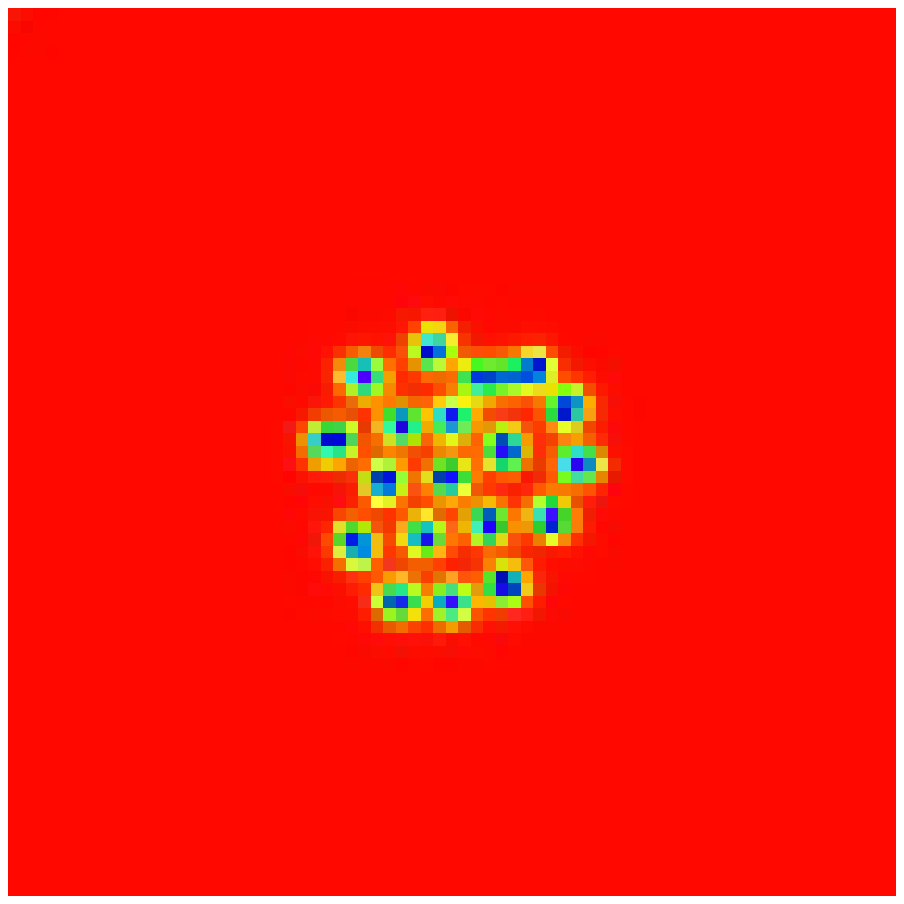} &
\includegraphics[width=0.24 \textwidth]{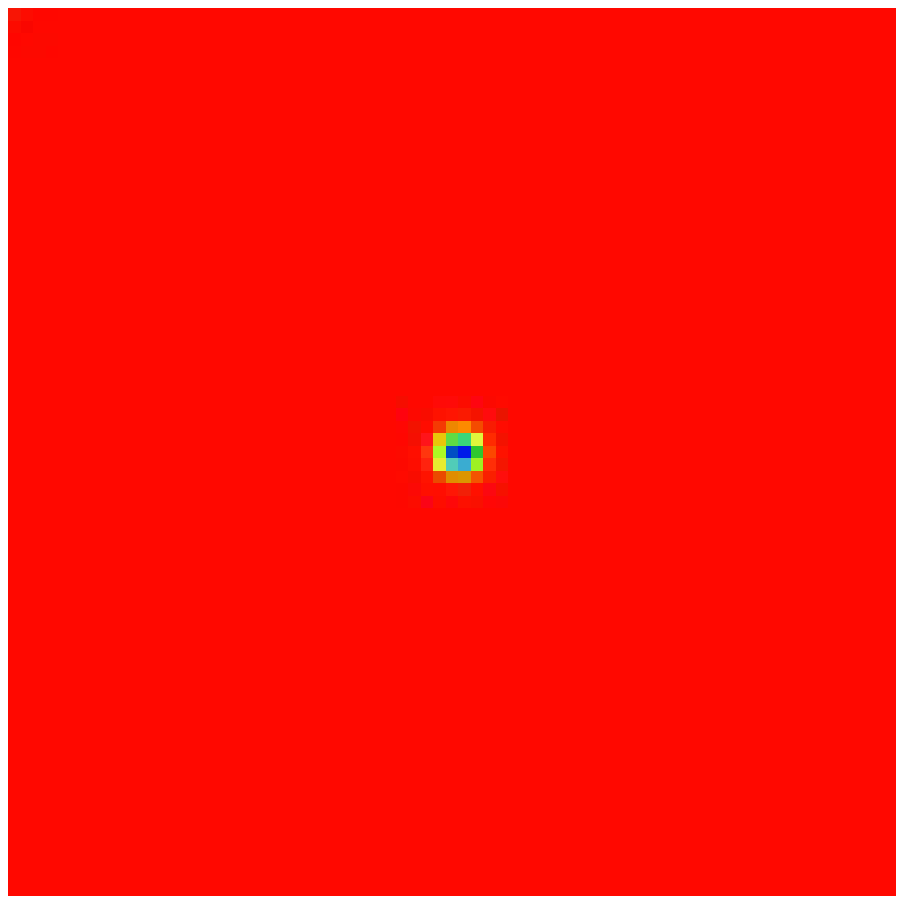}
\end{tabular}
\caption{\label{fig2}Results obtained with renormalized parameters
  $\lambda+\Delta \lambda$ and $\nu +\Delta \nu$ as given in Table
  \ref{table1}, and holding the weak noise fixed $A_v=4.8\times 10^{-6}$.
  Figures (a)-(d) are shown at $t=30000$.}
\end{figure}

We will now solve numerically the time evolution of this system (\ref{GSN})
holding all parameters \textit{fixed} and study the resulting
patterns as a function of increasing noise intensity $A_v$. First, for
$\lambda=1,D_v=0.5,D_u=1$, $u_0=0.05,\mu=0.1155,\nu=0.05$ and $A_u=0$, we
integrate this system with very weak noise $A_v=4.8\times 10^{-6}$
and obtain the pattern shown in Fig. \ref{fig1}-A. Then  we
increase only the noise intensity with the strength given in the
second column of Table \ref{table1} ($\lambda$ and $\nu$ are held fixed to the
values given in the next two columns) and obtain the patterns shown
in Figs. \ref{fig1}-B to D. The system evolves first forming long
stripes. As we vary only the noise, some of the stripes split
into spots and finally spots are the only stable structures. Noise
induces a transition between different patterns or states.

\begin{figure}
\begin{tabular}{ccc}
    (I) & (II) & (III)  \\
\includegraphics[width=0.24 \textwidth]{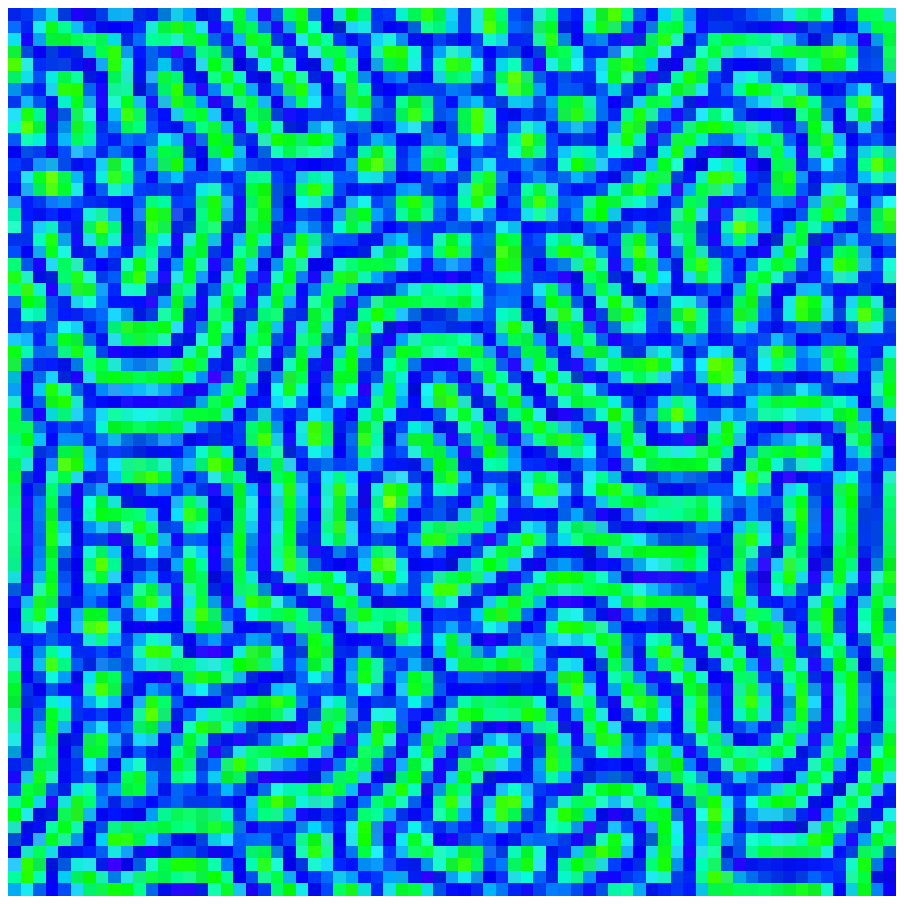} &
\includegraphics[width=0.24 \textwidth]{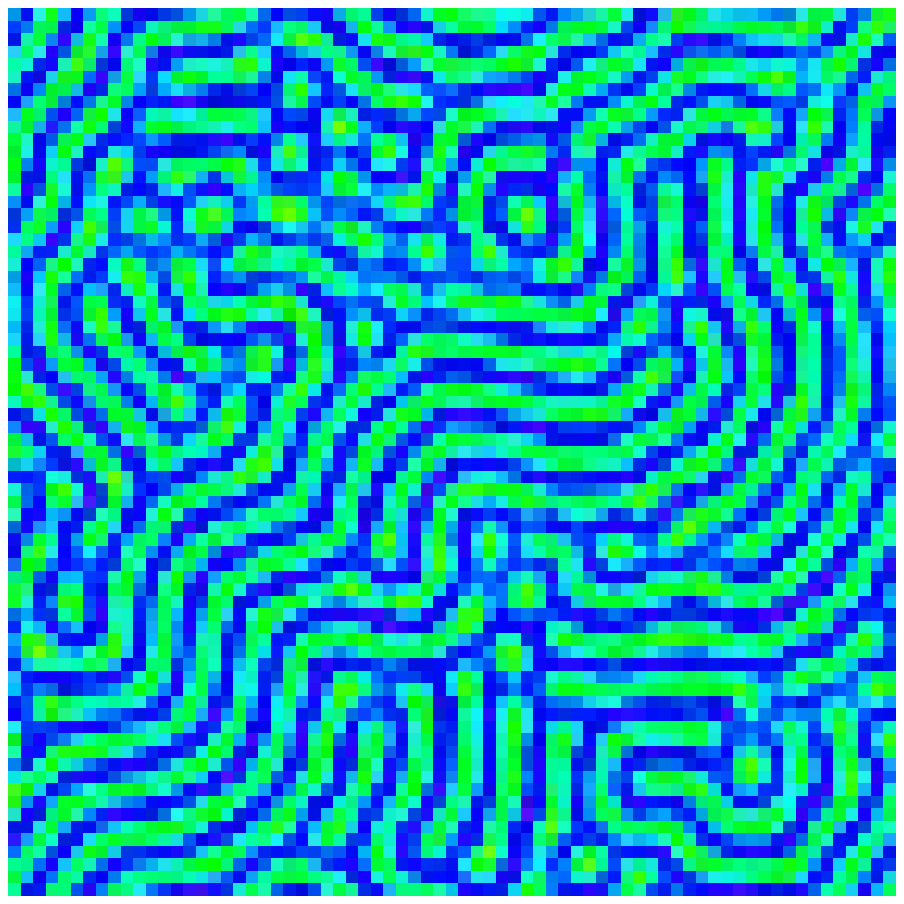} &
\includegraphics[width=0.24 \textwidth]{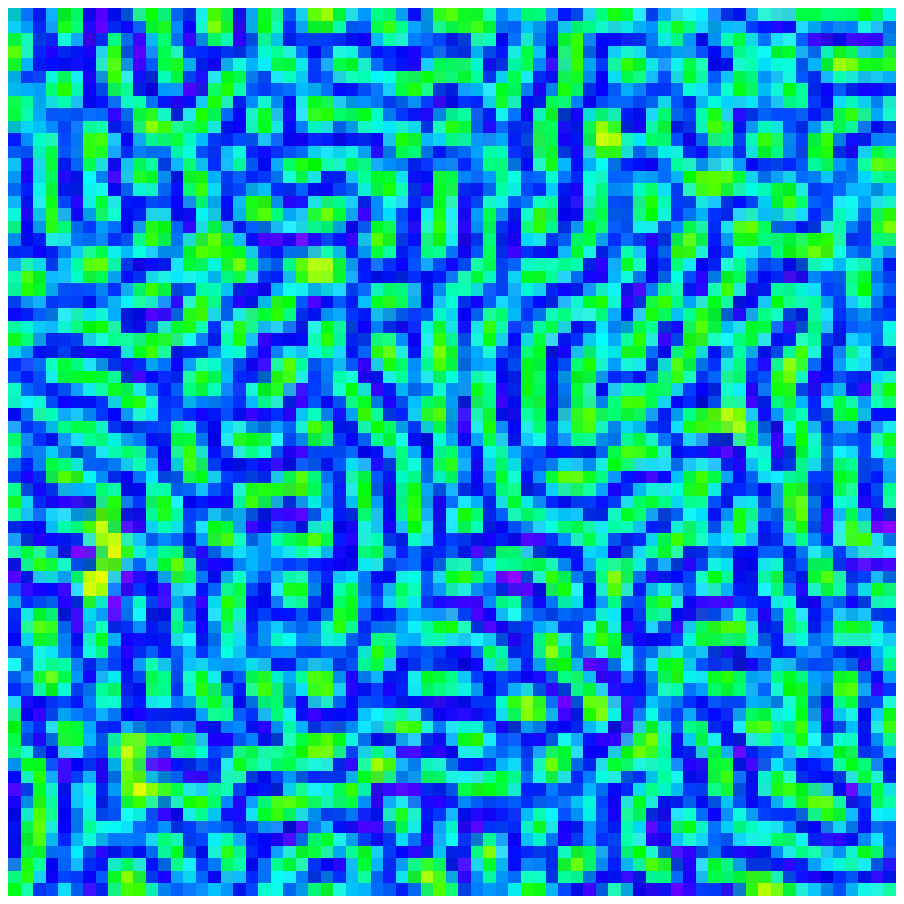}
\end{tabular}
\caption{\label{fig3}Results obtained with fixed parameters and increasing
  the noise intensity $A_v$ with the values given in column two of Table
  \ref{table2}. Figures (I)-(III) are shown at $t=20000$.  }
\begin{tabular}{ccc}
   (i) & (ii) & (iii) \\
\includegraphics[width=0.24 \textwidth]{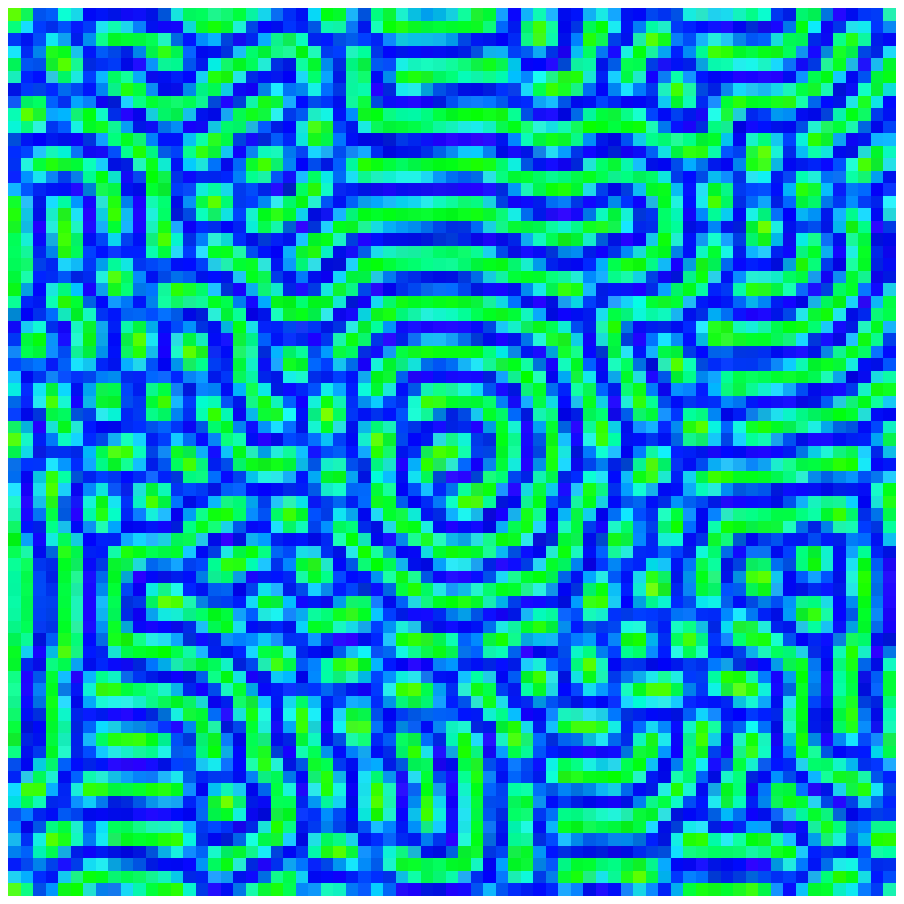}&
\includegraphics[width=0.24 \textwidth]{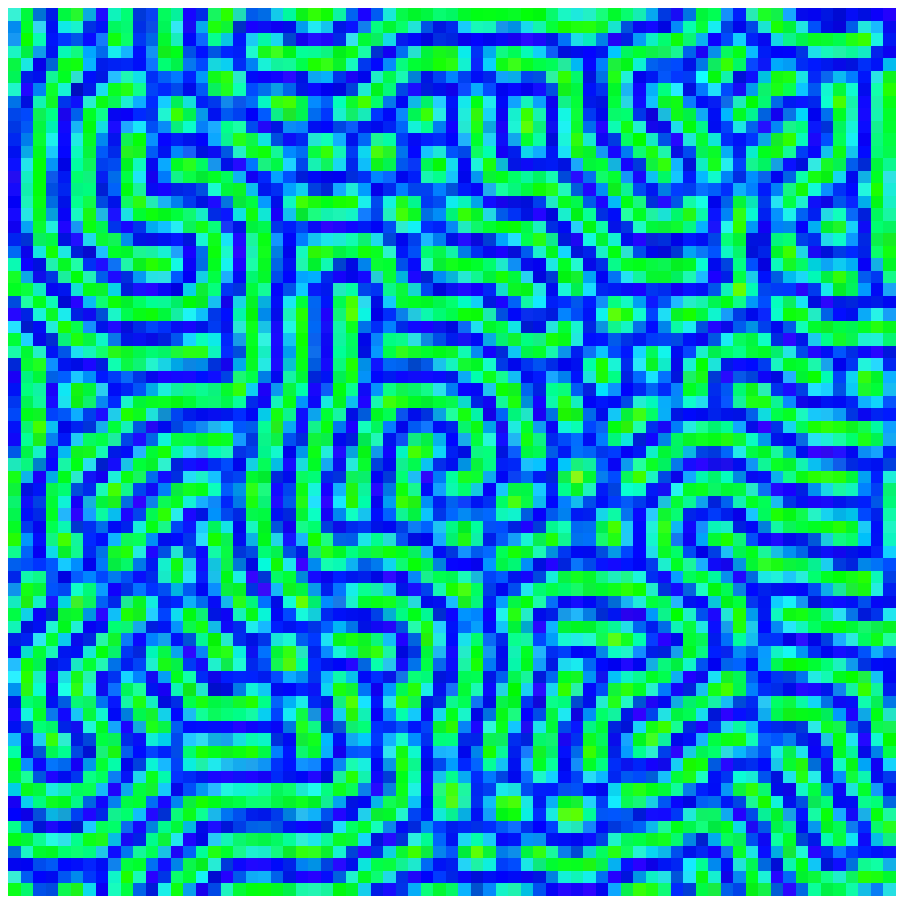}&
\includegraphics[width=0.24 \textwidth]{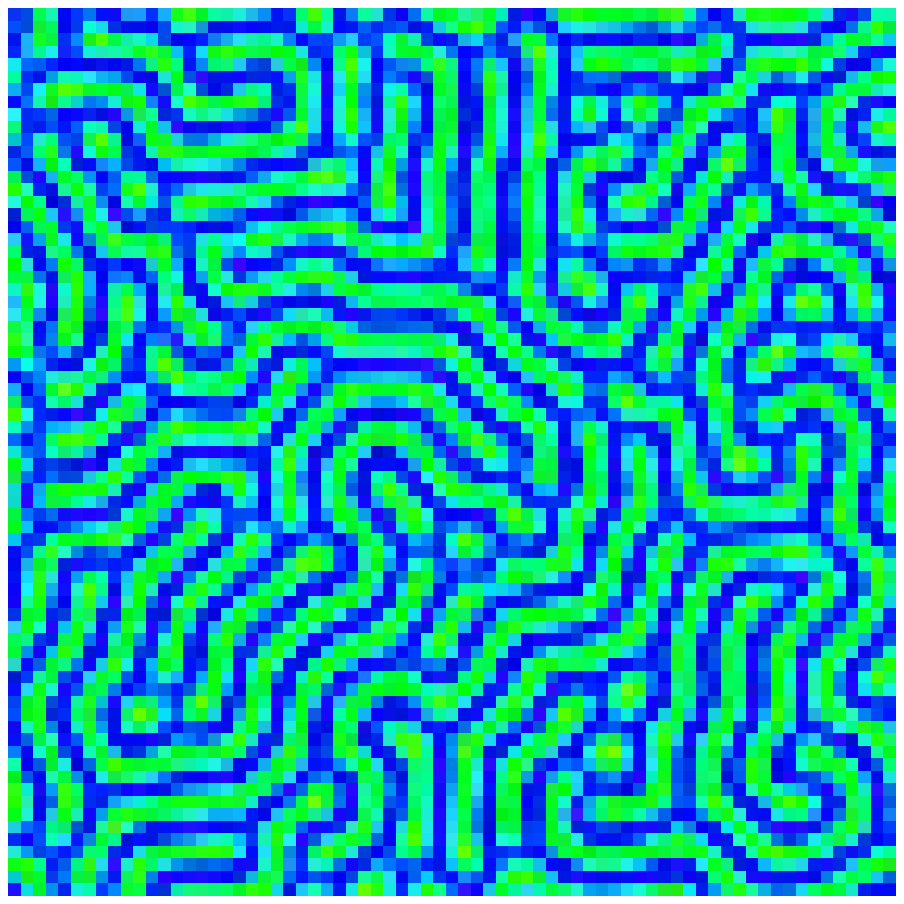}
\end{tabular}
\caption{\label{fig4}Results obtained with renormalized parameters
$\lambda+\Delta \lambda$ and $\nu +\Delta \nu$ as given in Table
\ref{table2}, and with a fixed weak noise $A_v=4.8\times 10^{-6}$.
Figures (i)-(iii) are shown at $t=20000$.}
\end{figure}

Next we repeat this study at a different location in parameter space:
$u_0=\nu=0.03,\mu=0.086$ keeping $\lambda=1,D_v=0.5,D_u=1$ and setting the
noise intensity  $A_v$ to the values given in the second column of
Table \ref{table2} ($\lambda$ and $\nu$ are held fixed to the
values given in the next two columns). We obtain now the patterns shown in
Figs. \ref{fig3}-I to III. For weak noise, the system fills in the space with a pattern of
short tubular structures and many spheres in clusters. Then, as we increase
the noise intensity, the spheres clusters
tend to disappear and the tubular structures become longer.

We have chosen two particular examples where weak noise induces an
observable transition between two patterns (in some other cases weak noise or
the equivalent small parameter variation does not
modify significatively the resulting pattern with respect to the reference
case).

\subsection{Transitions induced by parameter variation.}

Next we apply a fixed weak noise of intensity $A_v=4.8\times
10^{-6}$ and numerically solve the system (\ref{GSN}) with the
renormalized parameters, $\hat{\lambda}=\lambda+ \Delta \lambda$
and $\hat{\nu}=\nu+ \Delta \nu$, that correspond to the
coarse-grain scale $\Delta l = 2$ and the noise intensities
studied above which are given in the  sixth and seventh column of
Tables \ref{table1} and \ref{table2}. In Fig. \ref{fig2}-(a) to
(d), we show the results for $u_0=0.05$ and $\mu=0.1155$ and in
Fig. \ref{fig4}-(i) to (iii) for $u_0=0.03$ and $\mu=0.086$. The
results are equivalent to the patterns obtained before when only
the noise was applied. Within the limits of perturbation theory,
one can therefore obtain the same sequence of patterns and
morphologies either by increasing the noise intensity alone or by
holding the noise fixed and, applying a coarse-graining analysis,
 adjusting the $\lambda$ and $\nu$ parameters to their
renormalized values as predicted by the RG. This is the main
result of this paper.

Note that there is some discrepancy between the study cases
(d)-(D) and (iii)-(III): in case (d), where only the deterministic
parameters are changed, no self-replicating structure survives,
whereas in (D), with noise, the pattern persists and in case (iii)
the spheres have almost disappeared whereas in (III)  the structures are erased. This difference may indicate that we are
reaching the threshold where the higher order corrections (beyond one-loop) of the RG 
can no longer be ignored, or the point where the GS system is not renormalizable (new effective
reactions are generated which can not be described by a simple renormalization of the
original parameters as was explained in \cite{DH}).  We have observed
that generally, for $A_v\approx 4.5\times 10^{-4}$ and beyond, the
patterns can become too noisy and/ or non equivalent to the
patterns based on the renormalized parameter calculation. The dimensionless
perturbation parameter of the RG analysis performed in \cite{DH}
is $g=\frac{\lambda A_v K_2 \Lambda^{2}}{(\mu+ D_v \Lambda ^{2})}$
which when evaluated at this limit gives $g\approx 4.5 \times 10
^{-3}$.


Although the equivalence given by the RG equations breaks down at a certain
noise intensity, the change that a pattern will suffer when applying weak noise (for instance, from long
stripes into spots or from many spheres into long tubes) is easily predicted.

\section{Dynamic RG theory with finite scaling and correlated noise.}

Next, we include a study case of this system with temporally
correlated noise.  Following Eq. (\ref{RG_OU}) we observe, that
for a scale variation of $\Delta l$ the predicted modification of
the $\nu$ and $\lambda$ parameters induced by noise alone, should
be $\Delta \nu= + \Delta l
\csc{\frac{(1+2\theta_v)\pi}{2}}\frac{\lambda {A}_v K_2
\Lambda^{2}}{(\mu+ D_v \Lambda ^{2})^{1+2\theta_v}}$ and $\Delta
\lambda= - \Delta l \csc{\frac{(1-2\theta_v)\pi}{2}}\ \lambda
{A}_v K_2 \Lambda^{2} (\nu+ D_v \Lambda ^{2}) \frac{(\nu+ D_u
\Lambda ^{2})^{-1-2\theta_v}-(\mu+ D_v \Lambda
^{2})^{-1-2\theta_v}}{(\mu+ D_v \Lambda ^{2}) ^{2}-(\nu+ D_u
\Lambda ^{2}) ^{2}}\lambda$, respectively (where $\theta_v=1$).

\begin{figure}
\begin{tabular}{cccc}
    ($\alpha$) & ($\beta$) & ($\eta$) & ($\gamma$) \\
\includegraphics[width=0.24 \textwidth]{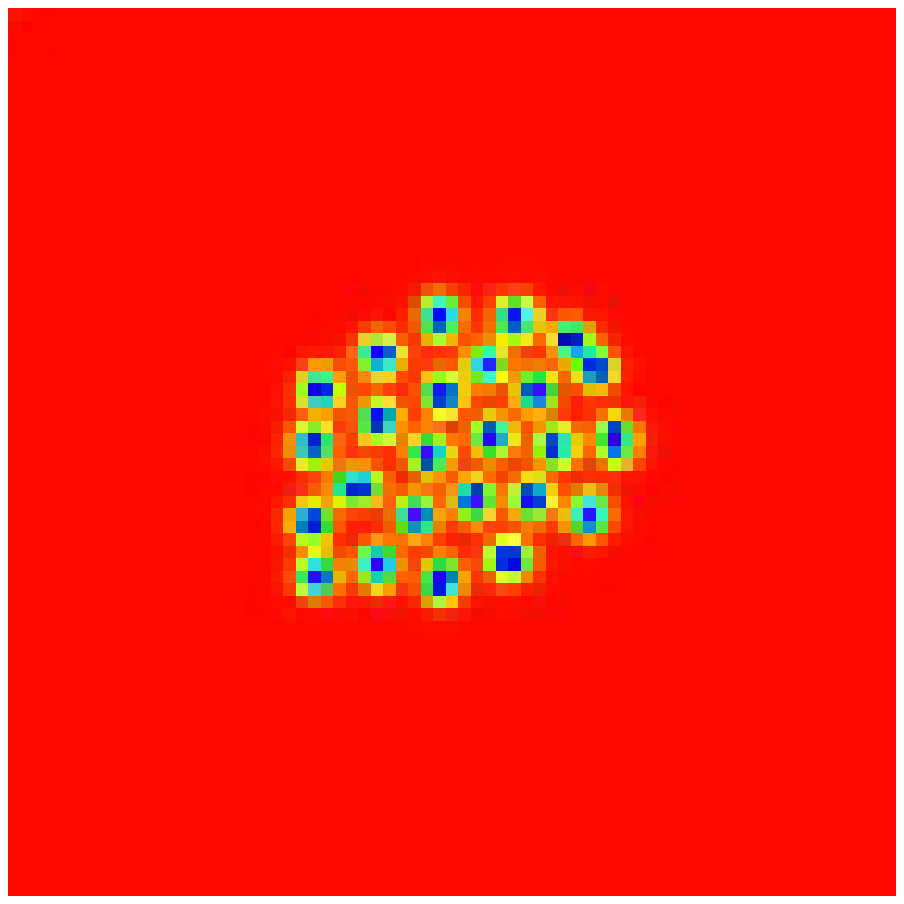} &
\includegraphics[width=0.24 \textwidth]{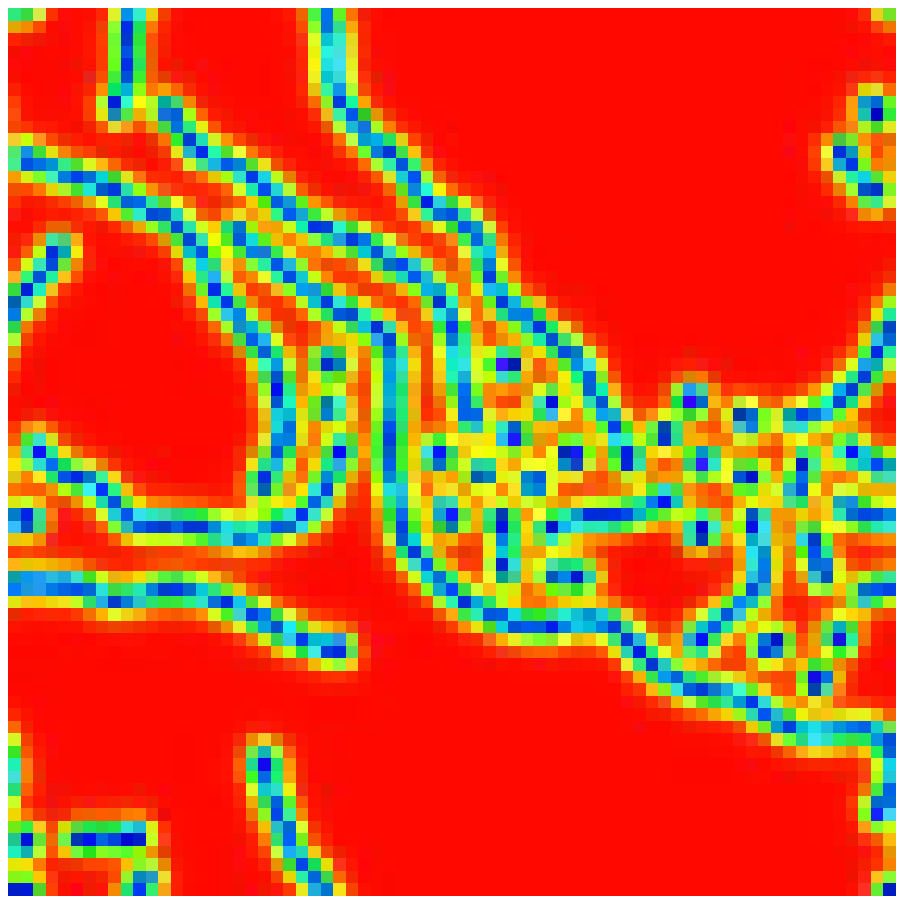} &
\includegraphics[width=0.24 \textwidth]{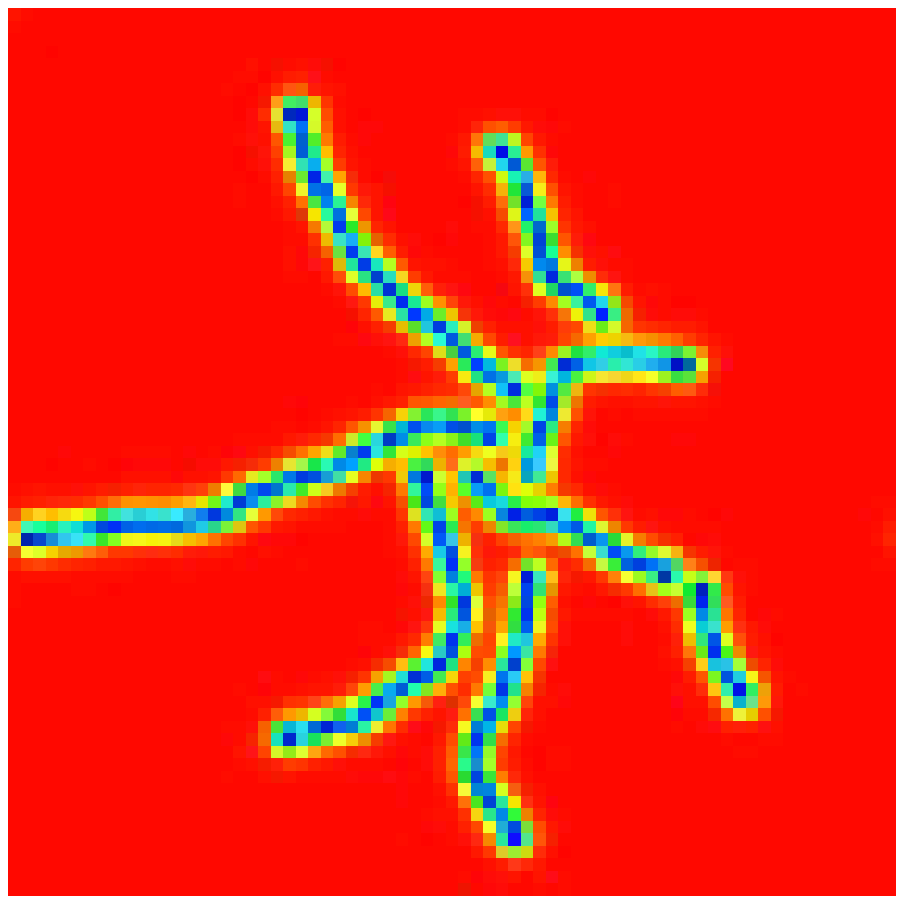} &
\includegraphics[width=0.24 \textwidth]{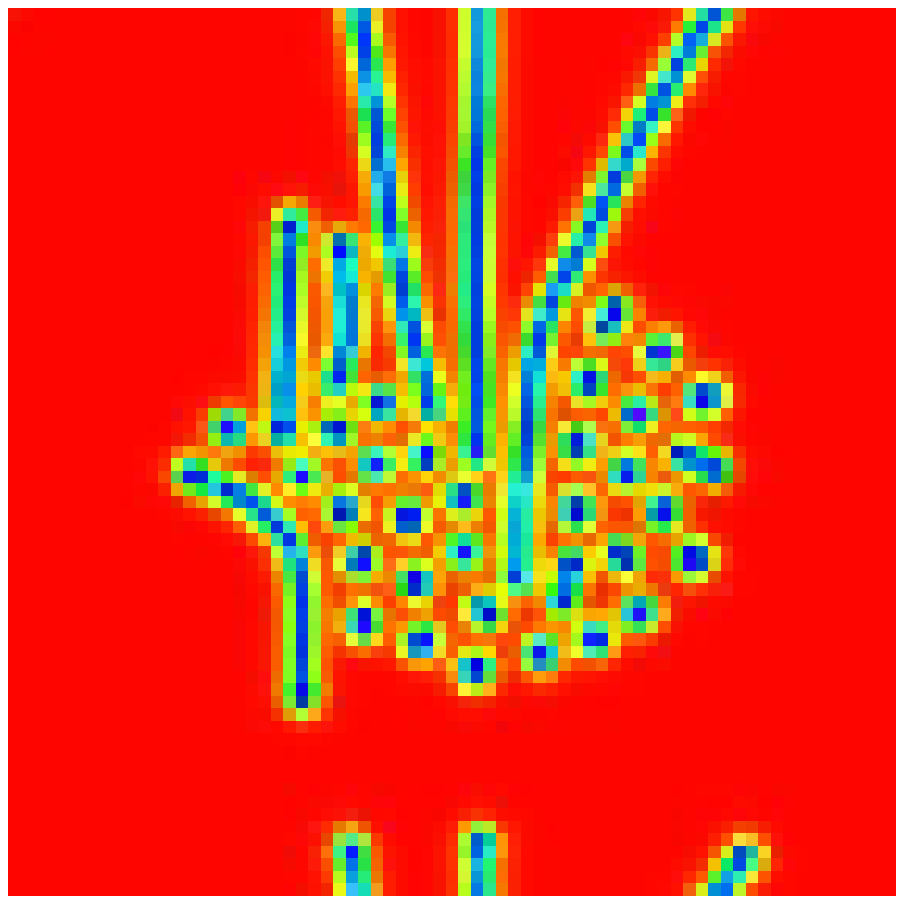}
\end{tabular}
\caption{\label{fig5} Results for correlated OU
  noise. ($\alpha$) is the starting point with $\mu=0.1155,u_0=0.05,\lambda=0.99991$ and $\nu=0.05027$
  ($\beta$) is
  obtained adding intermediate OU noise, ($\eta$) is
  obtained with stronger OU noise. Adding OU noise is equivalent to operating
  the system with new effective parameters: for the intermediate study case
  ($\beta$) this is shown in  ($\gamma$) which has been solved for $\hat{\lambda}=0.99996$ and
  $\hat{\nu}=0.05018$, for  case ($\eta$) the induced parameters are $\hat{\lambda}=1.0001$ and
  $\hat{\nu}=0.05$ as in the previously shown study case (a).  The patterns are given at $t=30000$.} 

\begin{tabular}{ccc}
     $(1)$ & $(2)$ & $(3)$\\
\includegraphics[width=0.24 \textwidth]{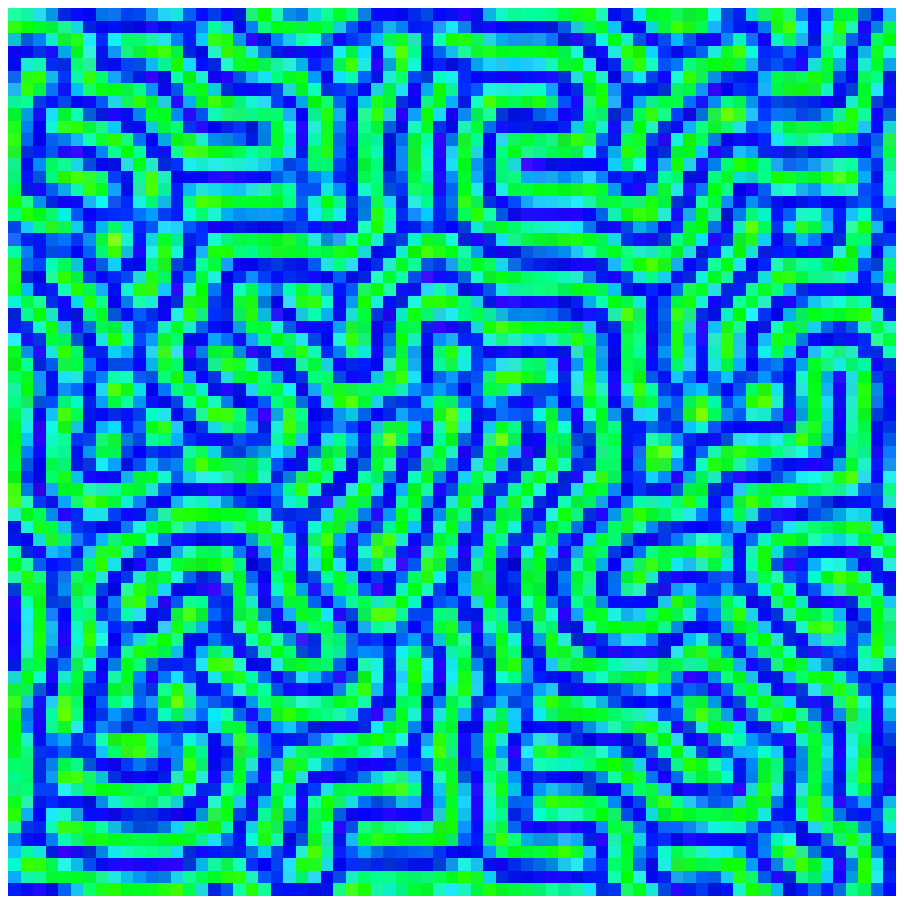} &
\includegraphics[width=0.24 \textwidth]{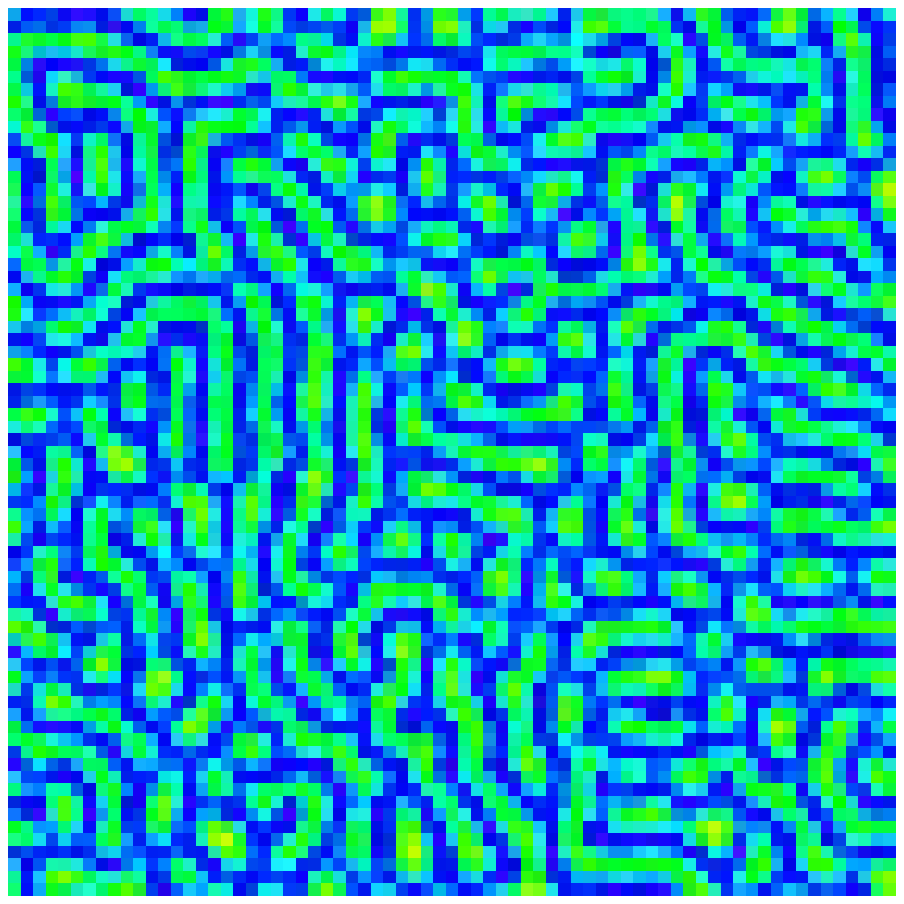} &
\includegraphics[width=0.24 \textwidth]{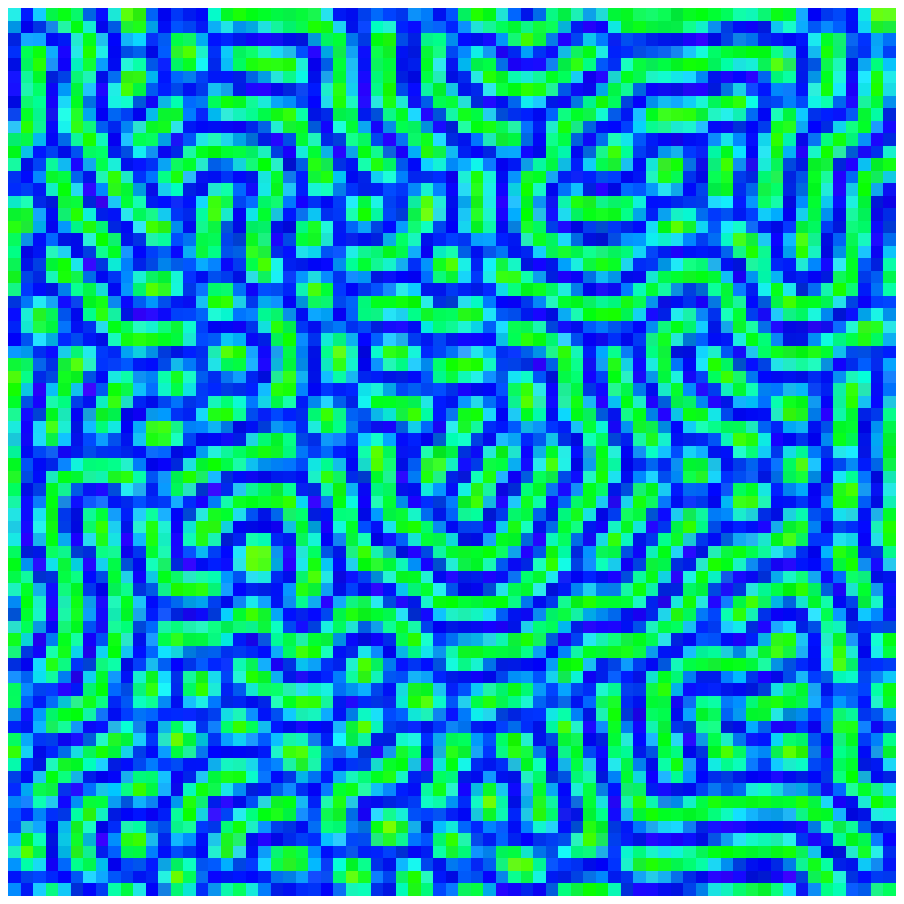}
\end{tabular}
\caption{\label{fig6} Results for correlated OU
  noise. ($1$) is the reference case for $\mu=0.086,u_0=0.03$,$\lambda=0.999914$ and $\nu=0.03026$
  ($2$) is
  obtained adding OU noise and ($3$) changing the system parameters
  to the new effective parameters $\hat{\lambda}=1.00001$ and $\hat{\nu}=0.030072$. The patterns are shown at $t=20000$.
 }
\end{figure}

 Our starting point now is equivalent to the previous study case (c), $\lambda=0.99991$ and $\nu=0.05027$, which (with the weakest white Gaussian noise $A_v=4.3 \times 10^{-4}$)
produces spots, see  Fig \ref{fig5}-($\alpha$). Next we add some OU noise
with ${A}_v=\hat{A}_v\tau=2.61\times 10^{-3}$ and $\tau=2000$, see Fig
\ref{fig5}-($\beta$), under this condition the system produces stripes and
spots. Increasing
the OU noise intensity further to ${A}_v=\hat{A}_v\tau=7.67\times 10^{-3}$ and
$\tau=5876$, see Fig
\ref{fig5}-($\eta$), the pattern
shows only long stripes. Adding OU noise of increasing strength to a system
operated with the
parameters of (c) we have reproduced the sequence (c)
to (a).

We can explain this behaviour as follows: this correlated noise has induced a
variation in $\lambda$ and $\nu$  of {\em opposite} sign to the case of white
Gaussian noise leaving the system with new effective parameters
 $\hat{\lambda}=1.0001$ and $\hat{\nu}=0.05$ for the study case ($\eta$).
If we solve the system with these renormalized parameters
we recover the same situation as in the (a) study case
which produces only stripes. For the intermediate case ($\beta$) the OU noise
has induced a variation to the new effective parameters
$\hat{\lambda}=0.99991$ and $\hat{\nu}=0.05027$. We have included for completeness a simulation
with these  values
(and the weakest white Gaussian noise $A_v=4.3\times10^{-4}$), see
Fig \ref{fig5}-($\gamma$), the resulting
pattern  with stripes and spots is equivalent to the ($\beta$) case as expected. 

Next we choose as starting point the
study case (iii), $\nu=0.03026$ and $\lambda=0.999914$, see Fig
\ref{fig6}-$(1)$, which produces long tubes and few spheres. We add some OU noise
with ${A}_v=\hat{A}_v\tau=5.22\times 10^{-3}$ and $\tau=2000$, see Fig
\ref{fig6}-$(2)$, under this condition the system produces more spheres and
shorter tubes. This
correlated noise induces a variation in the effective system parameters:
$\hat{\lambda}=1.00001$ and $\hat{\nu}=0.030072$. If we now solve the system
with the renormalized  parameters we obtain again
a pattern with more spheres and shorter tubes as in the study case $(2)$, see
Fig \ref{fig6}-$(3)$.

Applying correlated noise one can also find noise induced
transitions. The RG prediction for the parameter variation is also
valid for the case of correlated noise. Note that noise does not
always drive the system in the same pattern-sequence direction
since the sign of parameter variation/renormalization depends on
the type of noise.

\section{Conclusion}
We have studied the dynamics of the GS system with uncorrelated
and correlated noise. The environmental fluctuations, the spatial
diffusion and the non-linearities of the system create
cross-coupling terms and the new effective system parameters
depend on the noise amplitude and correlation time. We have shown
that for weak noises the lowest order one-loop RG analysis can be
applied at intermediate scale ranges, and be used to estimate the
effective parameters of the system. In other words, by combining
analytic and numerical work, we have established an equivalence
between a sequence of patterns generated by varying the noise
amplitude but keeping all other parameters fixed and a companion
sequence generated by keeping the noise fixed and varying (i.e.,
renormalizing) instead some of the model parameters according to
the RG flow equations.

For suitable noise
intensities the
perturbed values give rise to new patterns. Noise does not
always drive the system in the same pattern-sequence direction
since the sign of parameter variation depends on
the type of noise.
Based on this knowledge and using weak noise, one can lead the system
into a new state in a controlled way. 
 Notice also that adding noise to
a system  which is expected to produce structures of a certain spatial
correlation length (such as for instance the spots in the study case ($\alpha$))
does not necessary drive the system into the production of patterns with
shorter spatial correlations. On the contrary, it can
drive the dynamics of the system into a new situation which produces
patterns with longer spatial correlations (such as the stripes in the study case ($\eta$) with
OU noise).

This  raises important questions on the role of noise in chemical
and biological self-organization and the environmental selection of emergent
patterns. When describing a biological system not only should we estimate the
relevant parameters of the system, but also the magnitude of the stochastic
influences, the spatial scales involved and the correlation function. The
interdependence of most biological systems with the environment makes this
result specially relevant. In the
case of non-linear systems and in the vicinity of transition points, the
role of noise can be non-trivial, allowing the
system to explore new, well organized situations which might be favored by
the external natural selection processes. Applying the dynamic RG theory at
finite scales one can have an intuitive idea of how the system will be
driven by the environmental fluctuations.

\section{Acknowledgements}
We thank F. Lesmes for his involvement in the preliminary stages of the
numerical work. M.-P. Z. is supported by an Astrobiology fellowship from
the Instituto Nacional de T\'ecnica Aeroespacial (Spain). The work of D.H. and
F.M. is supported in part by MCyT (Spain) Grants \# BXX2000-1385 and \# BMC2000-0764, respectively.

\end{document}